\documentclass[twocolumn,secnumarabic,amssymb, nobibnotes, aps, prd,10pt,superscriptaddress]{revtex4}

\usepackage{graphicx}
\usepackage{float}
\usepackage[greek,english]{babel}
\usepackage{amsmath}

\newcommand{\unit}[1]{\ensuremath{\, \mathrm{#1}}}
\begin{document}
\newcommand{\gpi}{\textrm{\greektext p}}
\newcommand{\gmu}{\textrm{\greektext m}}

\title{Adsorption-induced modification of the hot electron lifetime in a Pb/Ag111 quantum well system}

\author{Florian Haag}
\email{f_haag@rhrk.uni-kl.de}
\affiliation{Department of Physics and Research Center OPTIMAS, TU Kaiserslautern, Erwin-Schroedinger-Straße 46, 67663 Kaiserslautern, Germany}
\affiliation{Graduate School of Excellence Materials Science in Mainz, Erwin-Schroedinger-Straße 46, 67663 Kaiserslautern, Germany}
\author{Tobias Eul}
\affiliation{Department of Physics and Research Center OPTIMAS, TU Kaiserslautern, Erwin-Schroedinger-Straße 46, 67663 Kaiserslautern, Germany}
\author{Lisa Grad}
\altaffiliation{Current address: Department of Physics, University of Zürich, Winterthurerstrasse 190, 8057 Zürich, Switzerland}
\author{Norman Haag}
\affiliation{Department of Physics and Research Center OPTIMAS, TU Kaiserslautern, Erwin-Schroedinger-Straße 46, 67663 Kaiserslautern, Germany}
\author{Johannes Knippertz}
\affiliation{Department of Physics and Research Center OPTIMAS, TU Kaiserslautern, Erwin-Schroedinger-Straße 46, 67663 Kaiserslautern, Germany}
\author{Mirko Cinchetti}
\affiliation{Experimentelle Physik VI, Technische Universität Dortmund, 44221 Dortmund, Germany}
\author{Martin Aeschlimann}
\affiliation{Department of Physics and Research Center OPTIMAS, TU Kaiserslautern, Erwin-Schroedinger-Straße 46, 67663 Kaiserslautern, Germany}
\author{Benjamin Stadtmüller}
\affiliation{Department of Physics and Research Center OPTIMAS, TU Kaiserslautern, Erwin-Schroedinger-Straße 46, 67663 Kaiserslautern, Germany}
\affiliation{Graduate School of Excellence Materials Science in Mainz, Erwin-Schroedinger-Straße 46, 67663 Kaiserslautern, Germany}

\begin{abstract}
  The interfacial band structures of multilayer systems play a crucial role for the ultrafast charge and spin carrier dynamics at interfaces.
  Here, we study the energy- and momentum-dependent quasiparticle lifetimes of excited states of a lead monolayer film on Ag(111) prior and after the adsorption of a monolayer of 3,4,9,10-perylene-tetracarboxylic-dianhydride (PTCDA).
  Using time-resolved two-photon momentum microscopy, we show that the electron dynamics of the bare Pb/Ag(111) bilayer system is dominated by isotropic intraband scattering processes within the quantum well state as well as interband scattering processes from the QWS into the Pb sideband.
  After the adsorption of PTCDA on the Pb monolayer, the interband scattering is suppressed and the electron dynamics is solely determined by intraband or inelastic scattering processes.
  Our findings hence uncover a new possibility to selectively tune and control scattering processes of quantum well systems by the adsorption of organic molecules.
\end{abstract}

\maketitle

\section{Introduction}

Interfaces between different types of functional materials are decisive building blocks for the next generation of nano-sized optoelectronic, photonic and spintronic applications \cite{Cinchetti.2017, Chiang.2000}.
They do not only determine the efficiency of charge and spin transport through the device structure but can also mediate device relevant functionalities, such as spin filtering or charge-to-spin and spin-to-charge conversion processes \cite{Sanchez.2013,Isasa.2016,Oyarzun.2016,Bergenti.2019}.
These functionalities are thereby closely linked to the interfacial band structure and the corresponding ultrafast carrier dynamics of the interfaces.

For this reason, countless studies focused on this correlation between the ultrafast single particle electron dynamics and the (spin-dependent) interfacial band structure of ultrathin metallic or molecular films on metallic surfaces \cite{Himpsel.1995,Bauer.1997,Ogawa.2002,Chulkov.2003,Wegner.2005,Kirchmann.2008,Zugarramurdi.2009,Hong.2009,Mathias.2010,Link.2000,ZHU.2004,Dutton.2005,Gudde.2006,Gundlach.2007,Hotzel.2007,Hagen.2010,Zugarramurdi.2012,Stadtmuller.2019}.
The most commonly used experimental technique for such studies is time-resolved two-photon-photoemission spectroscopy (tr-2PPE) \cite{Schmuttenmaer.1994,Wolf.1997,Petek.1997,Weinelt.2002,Bovensiepen.2012,Bauer.2015}, which allows one to characterize the ultrafast electron dynamics by the quasiparticle lifetime of the optically excited electrons in the so-called single-particle limit.
These studies laid the foundation for today’s understanding of the different energy and momentum dissipation mechanisms of (optically) excited charge and spin carriers at surfaces, interfaces and bulk materials.
For heterostructures and interfaces without structural order and well-defined bands, the quasiparticle lifetime only depends on the excited state energy of the excited electrons and is determined by inelastic electron-electron scattering processes \cite{Bauer.2015}.
Energy- and momentum-dependent quasiparticle lifetimes are in contrast only observed for sample systems with well-defined (interfacial) band structures \cite{Berthold.2002,Kirchmann.2008,Mathias.2010,Syed.2015,Monney.2016,Aeschlimann.2017,Ketterl.2018}.
Typical model systems with well-dependent band structures are e.g. quantum well states, image potential states or adsorption-induced shifted Shockley-surface states \cite{Kirchmann.2008,Weinelt.2007,Marks.2014}.
For these cases, electron-phonon or electron-defect scattering processes can lead to coupled energy and momentum dissipation processes of excited electrons, which are typically classified as inter- or intraband scattering processes.
These scattering processes can be directly identified in a time- and angle-resolved 2PPE experiment due to the strong correlation between the energy- and momentum-dependent quasiparticle lifetimes of optically excited electrons and the band structure of metal-metal or metal-organic interfaces.

To gain further insights into the inter- or intraband scattering processes at interfaces, we turn to the single-particle electron dynamics of one layer lead (Pb) on an Ag(111) single crystal surface.
This model system was selected as an exemplary case from the manifold of ultrathin layers on surfaces that can host quantum confined electrons and quantum well states (QWSs) \cite{Ogawa.2002,Wegner.2005,Kirchmann.2008,Mathias.2010,Stadtmuller.2019b}.
In the particular case of the Pb/Ag(111) interface, the band structure reveals two distinct bands in the Pb layer: (i) a parabolic, free electron-like QWS with parabolic dispersion in the center of the surface Brillouin zone and p$_{\unit{z}}$ orbital character and (ii) a Pb side band with almost linear dispersion and p$_{x/y}$ orbital character \cite{Stadtmuller.2019b}.
These two bands are expected to dominate the ultrafast electron dynamics of such ultrathin Pb films.

Previous tr-2PPE studies of similar ultrathin Pb layers already proposed complex inter- and intraband scattering processes between diﬀerent Pb and Pb-derived states \cite{Kirchmann.2008,Mathias.2010} in a limited part of the surface Brilliouin zone (either conducted without angular (momentum) resolution or along one high symmetry direction in momentum space).
We build on these pioneering studies using time-resolved two-photon momentum microscopy (tr-2PMM) \cite{Haag.2019}.
This approach combines the optical setup of a conventional tr-2PPE experiment with a photoemission electron microscope operated in momentum space mode, i.e., a momentum microscope \cite{Escher.2005,Kromker.2008,Tusche.2015,Schonhense.2015}, which allows us to study inelastic as well as \mbox{(quasi-)}elastic interband and intraband scattering processes in the whole accessible momentum space with a parallel detection scheme and a fixed experimental geometry.
In a second step, we explore the tuneability of the electron dynamics of the Pb/Ag(111) bilayer system by adding the prototypical molecule 3,4,9,10-perylene-tetracarboxylic-dianhydride (PTCDA) on top of the Pb/Ag(111).
The adsorption of PTCDA alters the band structure of the Pb/Ag interface thereby suppressing one of the prominent scattering channels.

Our tr-2PMM experiments allow us to disentangle the intrinsic quasiparticle lifetime of the Pb side band at the bare Pb/Ag(111) interface.
We find that the electron dynamics of the Pb/Ag(111) bilayer system is dominated by isotropic scattering processes of electrons of the Pb QWS in momentum space.
They lead to a lifetime of $\tau\approx 12\,\unit{fs}$, which is isotropic in momentum and increases slightly when approaching the band minimum due to intraband scattering.
Additionally, we observe isotropic interband scattering of electrons from the QWS into the Pb side band with p$_{\unit{x/y}}$ orbital character.
This scattering process is responsible for an increased lifetime in the Pb side band despite its intrinsically vanishing quasiparticle lifetime.
After the adsorption of PTCDA on the Pb monolayer, all interband scattering processes are suppressed and the electron dynamics is solely determined by intraband or inelastic scattering processes.
In this way, our study demonstrates an alternative way for tuning and controlling scattering processes in QWS systems by the adsorption of organic molecules.

\section{Experimental Details}
\subsection{Sample Preparation}
  All experiments and the sample preparation procedures were performed under ultrahigh-vacuum conditions with a base pressure better than $1\cdot10^{-9}$ mbar.
  The surface of the (111)-oriented silver crystal was cleaned by several cycles of argon ion bombardments and subsequent annealing at a temperature of $T_{\unit{Sample}}=730\,\unit{K}$.
  The cleanliness of the Ag(111) surface was verified by the width of the diffraction spots in low energy electron diffraction (LEED) measurements and the linewidth of the Shockley surface state at the $\overline{\Gamma} $-point using two-photon photoemission momentum microscopy.
  Afterwards, more than a ML Pb was evaporated onto the substrate at room temperature.
  Subsequent annealing for 10 minutes up to 420 K leads to desorption of higher Pb layers resulting in a uniform single Pb layer.
  The result was checked with LEED and momentum-resolved photoemission spectroscopy.
  The coverage of the molecule PTCDA was controlled by the evaporation time and the flux of the evaporator.

\subsection{Time-resolved two-photon momentum microscopy and analysis}

  The time-resolved two-photon momentum microscopy measurements (tr-2PMM) were performed with a photoemission electron microscope (PEEM) operated in k-space mode.
  We used a time-of-flight detector as energy analyzer \cite{Oelsner.2001,Oelsner.2010,Schonhense.2015} for the Pb ML measurements and a double hemispherical analyzer \cite{Escher.2005,Kromker.2008,Tusche.2015} for the one layer PTCDA on Pb/Ag(111) system.
  Both microscopes are combined with optical beamlines for pump-probe spectroscopy \cite{Haag.2019}.
  As light sources for the optical part of our experiments, we used the second harmonics (SHG) of titanium-sapphire laser oscillators with a central wavelength of $800\,\unit{nm}$ ($1.55\,\unit{eV}$) for the fundamental emission line, sub $30\,\unit{fs}$ pulse width (FWHM), and a repetition rate of $80\,\unit{MHz}$.
  The polarization was changed within the experiments using a $\lambda$-half-wave plate. A mirror located within the PEEM optics was used for the nearly normal incidence angle measurements (4° incidence angle with respect to the surface normal) \cite{Kahl.2014}.

  The multi-dimensional data sets $I(k_{\unit{x}},k_{\unit{y}},$E-E$_{\unit{F}},\Delta t)$ were analyzed using the same approach as described in our previous work \cite{Haag.2019}.
  The momentum-dependent lifetimes were determined by extracting autocorrelation curves at each point in the momentum space for each intermediate state energy E-E$_{\unit{F}}$.
  These traces were analyzed within the framework of optical Bloch equations which yields lifetimes in momentum space \cite{Aeschlimann.1996,Hertel.1996,Ogawa.1997}.
  These lifetimes are plotted as color code in so-called lifetime maps, which reflect the lifetimes of electrons throughout the entire accessible momentum space in a tr-2PMM experiment.
  Note that these lifetimes do not necessary reflect the pure intrinsic quasiparticle lifetime of the corresponding state in energy and momentum space but also contains signatures of energy- and momentum-dependent refilling processes.

\section{Results and Discussion}
\begin{figure}
     \includegraphics[width=65mm]{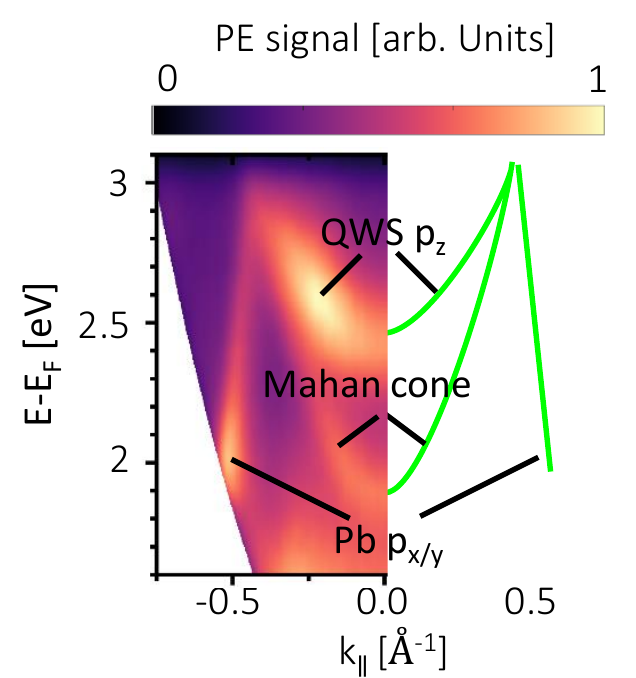}
  \caption{Left: Intermediate state energy E-E$_{\unit{F}}$ over $k_{\parallel}$ cut along the $\overline{\unit{M}}'\,\overline{\Gamma}$-direction.
  Right: Sketch of the $\overline{\Gamma}\,\overline{\unit{M}}$-direction.
  The parabolic features refer to the Pb QWS having a p$_{\unit{z}}$ orbital character and the hybridized sp-sp Mahan cone transition.
  The straight lines belong to the Pb side band, having a p$_{\unit{x/y}}$ orbital character.
  All bands are named accordingly.}
  \label{fig:Bild1}
  \end{figure}

  We start our discussion with the hot electron dynamics of the bare Pb/Ag(111) bilayer system.
  The unoccupied band structure of the material system is illustrated in \mbox{Fig.\ref{fig:Bild1}} as an energy vs. momentum cut along the $\overline{\unit{M}}'\,\overline{\Gamma}$-direction.
  Here, we only review the most important spectroscopic features of the unoccupied band structure which are sketched on the right side of \mbox{Fig.\ref{fig:Bild1}}.
  A more detailed discussion can be found elsewhere \cite{Stadtmuller.2019b}.

  The unoccupied band structure of the Pb/Ag(111) bilayer system is dominated by two Pb-derived states: (i) a free electron-like quantum well state (labelled QWS p$_{\unit{z}}$) with p$_{\unit{z}}$ orbital character and a parabolic dispersion centred at the $\overline{\Gamma}$-point of the surface Brillouin zone and (ii) a Pb-derived side band (labelled Pb p$_{\unit{x/y}}$) with almost linear dispersion and p$_{\unit{x/y}}$ orbital character.
  The third spectroscopic feature with parabolic dispersion (labelled Mahan cone) is not a band of the unoccupied Pb/Ag band structure, but can be attributed to a so-called Mahan Cone transition \cite{Mahan.1970,Winkelmann.2012}.
  It is caused by a resonant transition of electrons from an occupied valence band into an unoccupied band above the vacuum energy via a strongly detuned intermediate state (virtual intermediate state).
  For the Pb/Ag(111) system, this transition only becomes visible for exciting light with a non-vanishing out-of-plane electric field vector.

\begin{figure*}
  \centering
     \includegraphics[width=160mm]{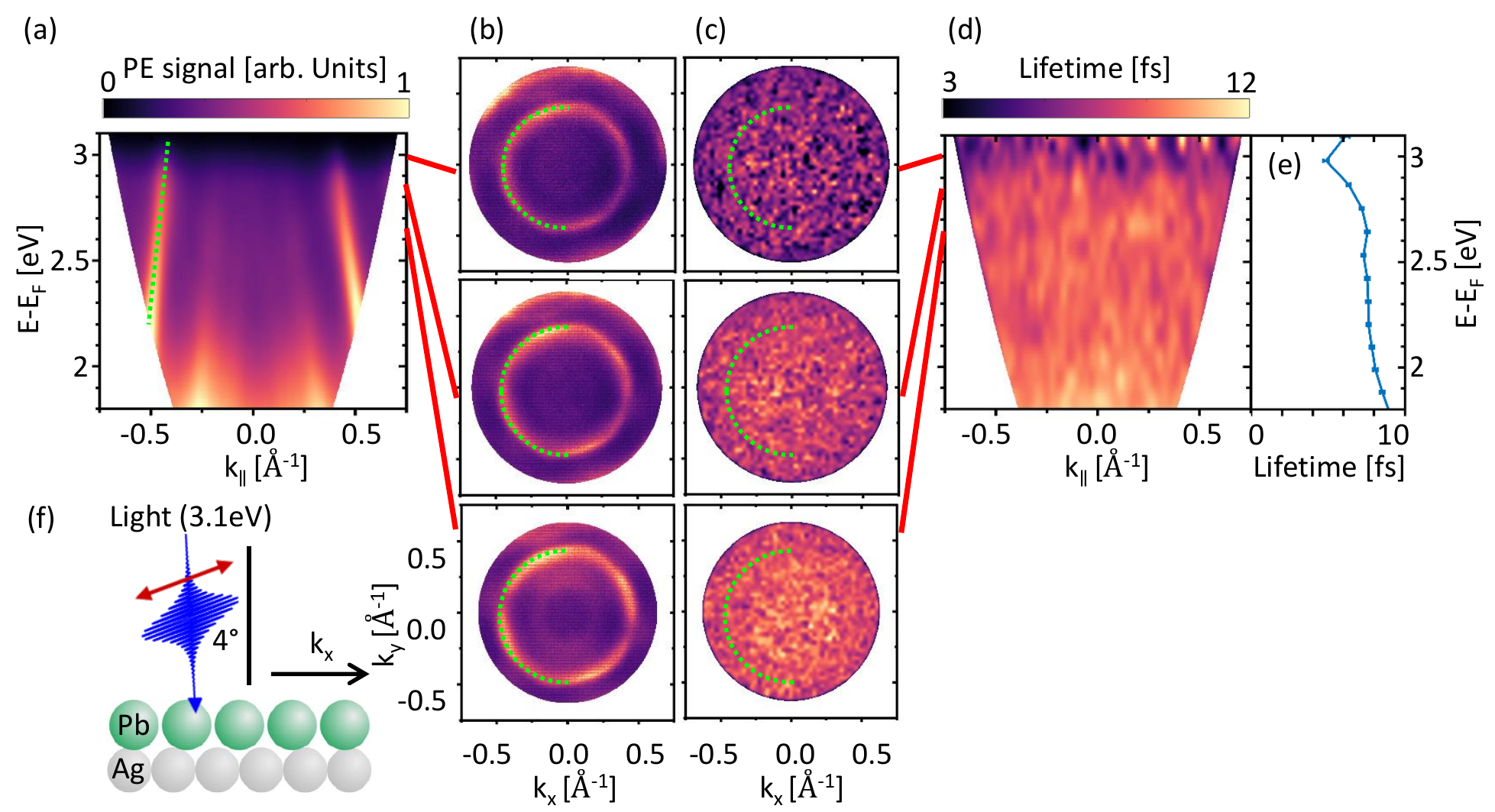}
  \caption{Time-resolved two-photon momentum microscopy experiment of the Pb ML on Ag(111) system performed with an incidence angle of 4° (nearly normal incidence, in-plane electric field vector parallel to the $k_{\unit{x}}$-direction) and a photon energy of $3.1\,\unit{eV}$.
  (a) E over $k_{\parallel}$ cut along the $\overline{\unit{M}}'\,\overline{\Gamma}\,\overline{\unit{M}}$-direction.
  (b) Constant energy maps for E-E$_{\unit{F}}=2.65\,$eV, $2.85\,$eV and $3.0\,$eV starting from the bottom to the top.
  (c) Constant energy momentum-dependent lifetime maps corresponding to the photoemission images shown in (b).
  (d) Corresponding lifetime map to the two-photon photoemission signal shown in (a).
  The band dispersion of the Pb side band is indicated with green dotted lines as a guide-to-the-eye.
  (e) Energy-dependent momentum-integrated quasiparticle lifetime.
  (f) Experimental geometry, photon energy and sample system of the experiment.}
  \label{fig:Bild2}
  \end{figure*}

  The energy- and momentum-dependent electron dynamics of the Pb-derived side band can be extracted from the tr-2PMM data set shown in \mbox{Fig.\ref{fig:Bild2}}.
  These data were obtained in a monochromatic ($3.1\,\unit{eV}$) 2PPE experiment in nearly normal incidence geometry with an in-plane electric field vector parallel to the $k_{\unit{x}}$-direction (see sketch in \mbox{Fig.\ref{fig:Bild2}(f)}).
  In this experimental geometry, the light pulses of the pump- and probe beam only exhibit an electric field component parallel to the surface and hence can only excite and probe states with predominant in-plane orbital character.
  This is clearly reflected in the energy vs. momentum cut along the $\overline{\unit{M}}'\,\overline{\Gamma}\,\overline{\unit{M}}$-direction in \mbox{Fig.\ref{fig:Bild2}(a)}, which only shows the spectroscopic signature of the Pb side band with p$_{\unit{x/y}}$ orbital character (marked by a green dotted line).
  The dispersion of this state in momentum space is reflected in the exemplary constant energy (CE) maps in \mbox{Fig.\ref{fig:Bild2}(b)} which were recorded at three intermediate state energies of E-E$_{\unit{F}}=2.65\,$eV, $2.85\,$eV and $3.0\,$eV.
  In all CE maps, this state appears as a ring-like shape with an azimuthal intensity distribution.
  The intensity modulation is the result of the in-plane orbital character and the p3m1 symmetry of the Pb superstructure on Ag(111).

  The results of the tr-2PMM experiment are shown in different representations in \mbox{Fig.\ref{fig:Bild2}}(c)-(e).
  The momentum-dependent quasiparticle lifetime is displayed in three exemplary lifetime maps in \mbox{Fig.\ref{fig:Bild2}(c)}, which were obtained at the same intermediate state energies at the momentum-resolved emission patterns in \mbox{Fig.\ref{fig:Bild2}(b)}.
  The color code of these maps represents the quasiparticle lifetime at each position in momentum space.
  It was obtained by an individual fit of the momentum-dependent autocorrelation trace using optical Bloch equations (see \cite{Haag.2019} for more details).

  All lifetime maps show momentum-independent quasiparticle lifetimes of the electrons of only a few femtoseconds, which slightly increases when reducing the intermediate state energy towards the Fermi energy.
  In particular, the CE lifetime maps do not show any momentum space pattern that resembles the dispersion of the Pb side band.
  The latter is marked in the lifetime maps by green half circles.
  This observation indicates a neglectable intrinsic quasiparticle lifetime of the Pb side bands, i.e., the lifetime is indistinguishable from the lifetime of the homogeneous background in the L-bandgap of the Ag(111) crystal.

  These findings are also reflected in the energy vs. momentum lifetime map along the $\overline{\unit{M}}'\,\overline{\Gamma}\,\overline{\unit{M}}$-direction in \mbox{Fig.\ref{fig:Bild2}(d)}, which does not show any pattern following the linear dispersion of the Pb side bands.
  The energy-dependent and momentum averaged quasiparticle lifetime is quantified in \mbox{Fig.\ref{fig:Bild2}(e)}, which reveals an increase from $\tau\approx 6\,$fs at E-E$_{\unit{F}}=3.0\,\unit{eV}$ to $\tau\approx 9\,\unit{fs}$ at E-E$_{\unit{F}}=1.8\,\unit{eV}$.

\begin{figure*}
  \centering
     \includegraphics[width=148mm]{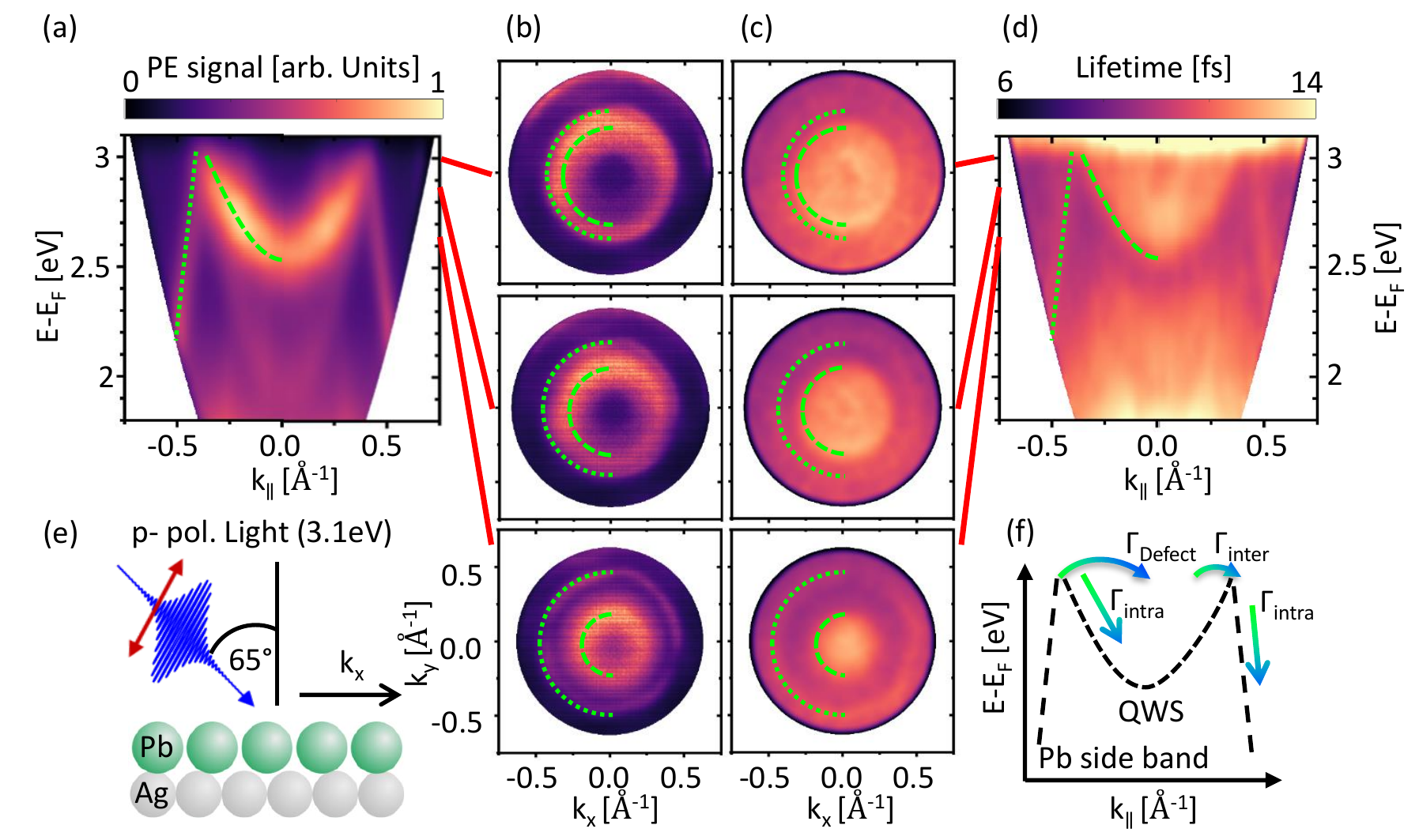}
  \caption{Time-resolved two-photon momentum microscopy experiment of the Pb ML on Ag(111) system performed with p-polarized light, a gracing incidence angle of 65° and a photon energy of 3.1 eV.
  (a) E over $k_{\parallel}$ cut along the $\overline{\unit{M}}'\,\overline{\Gamma}\,\overline{\unit{M}}$-direction.
  (b) Constant energy maps for E-E$_{\unit{F}}=2.65\,$eV, $2.85\,$eV and $3.0\,$eV starting from the bottom to the top.
  (c) Constant energy momentum-dependent lifetime maps corresponding to the photoemission images shown in (b).
  (d) Corresponding lifetime map to the photoemission signal shown in (a).
  The band dispersion of the Pb quantum well state (green dashed line) and the Pb side band (green dotted line) are indicated as a guide-to-the-eye.
  (e) Experimental geometry, photon energy and sample system of the experiment.
  (f) Exemplary sketch of the band structure with the occurring scattering processes.}
  \label{fig:Bild3}
  \end{figure*}

  Next, we turn to our tr-2PMM data set obtained for p-polarized light and a gracing incidence angle of $65^\circ$ with respect to the surface normal (the in-plane component of the electric field vector oscillates parallel to the $k_{x}$-direction).
  The results are summarized in \mbox{Fig.\ref{fig:Bild3}}.
  The energy vs. momentum cut through the unoccupied band structure along the $\overline{\unit{M}}'\,\overline{\Gamma}\,\overline{\unit{M}}$-direction in \mbox{Fig.\ref{fig:Bild3}(a)} reveals spectroscopic signatures of the QWS, the Pb side band as well as of the Mahan cone transition.
  All Pb-derived states are now accessible in our 2PMM experiment due to the out-of-plane component of the electronic field vector of the pump and probe pulses in gracing incidence geometry (see sketch in \mbox{Fig.\ref{fig:Bild3}(e)}).
  Again, the dispersions of all states are shown in the exemplary constant energy (CE) maps in \mbox{Fig.\ref{fig:Bild3}(b)}, which were recorded at three intermediate state energies E-E$_{\unit{F}}=2.65\,$eV, $2.85\,$eV and $3.0\,$eV.
  In analogy to the 2PMM experiment in normal emission geometry, the Pb side band reveals a ring-like shape with azimuthal intensity distribution.
  This intensity variation is visible most clearly in the CE map at E-E$_{\unit{F}}=2.65\,$eV (bottom CE map of panel (b)).
  In contrast, the predominant p$_{\unit{z}}$ orbital character of the Pb QWS and the states involved in the Mahan cone transition results in ring-like emission pattern of these features with a homogeneous azimuthal intensity distribution.

  The lifetimes of these states are reflected in the CE lifetime maps in \mbox{Fig.\ref{fig:Bild3}(c)}.
  These CE maps reveal a disc-like and a ring-like feature with quasiparticles lifetimes that are distinguishable from those of the homogeneous background in the L-bandgap of the Ag(111) crystal.
  Both features follow the dispersion of the QWS (dashed circles) as well as of the Pb side band (dotted circles) closely in all CE maps and hence allow us to gain insights into the momentum-dependent lifetimes of both states.

  The disc-like feature reflects the lifetimes of the QWS in momentum space.
  Its radius in momentum space follows very closely the energy-dependent band dispersion of the QWS as indicated by the dashed green circles in the CE lifetime maps in \mbox{Fig.\ref{fig:Bild3}(c)}.
  This is even more clearly visible in the energy vs. momentum lifetime map in \mbox{Fig.\ref{fig:Bild3}(d)}, where the energy and momentum regions with distinct lifetime resembles closely the parabolic dispersion of the QWS (green dashed curve).
  Along this parabolic dispersion, we find slight increase of the lifetime of the QWS from $\approx 12.0\,\unit{fs}$ at E-E$_{\unit{F}}=3.0\,\unit{eV}$ to $\approx 13.2\,$fs at E-E$_{\unit{F}}=2.7\,\unit{eV}$.
  Focusing on the momentum-dependent lifetimes of the QWS, we do not find any azimuthal variation of the quasiparticle lifetime for all energies.
  However, the disk-like pattern of the QWS itself indicates the existence of an additional (quasi-)elastic, but isotropic scattering process of electrons from the QWS with ring-like dispersion in momentum space towards the $\overline{\Gamma}$-point of the Brillouin zone.

  The second feature with a distinct lifetime in the CE lifetime maps exhibits a ring-like pattern with increasing radius for decreasing intermediate state energies.
  This energy-dependent radii resembles perfectly the dispersion of the Pb side band with p$_{\unit{x/y}}$-orbital character (see green dotted line) as shown in the energy vs. momentum lifetime map in \mbox{Fig.\ref{fig:Bild3}(d)}.
  The lifetime of this feature is $\tau\approx 11\,\unit{fs}$ for all energies and azimuthal orientations.
  This value is significantly larger than the lifetimes of the Pb side band obtained in normal incidence geometry and hence does not reflect the intrinsic quasiparticle lifetime of this state.
  Instead, we propose that this apparently larger lifetime is due to a momentum-dependent refilling of electrons from the QWS to the Pb side band mediated by a interband scattering process.
  Such a process was already proposed for the QWS system Pb/Cu(111) \cite{Mathias.2010}.

  Altogether, the electron dynamics and the different inter- and intraband scattering processes are summarized in \mbox{Fig.\ref{fig:Bild3}(f)}.
  Optically excited electrons in the QWS dissipate energy and momentum by isotropic intraband scattering following the band dispersion towards the band bottom of the QWS.
  Additional (quasi-)elastic scattering processes can isotropically redistribute electrons from the QWS either towards the center of the Brillouin zone, most likely by electron-defect scattering, or into the Pb side bands via interband scattering between both Pb-derived bands.
  Most importantly, only the interband scattering process from the QWS into the Pb side band leads to an increased lifetime of the Pb side band which otherwise reveals a vanishing intrinsic quasiparticle lifetime for all energies and momenta addressed in our experiment.

  In the next step, we focus on the modifications of the ultrafast electron dynamics of the Pb/Ag(111) bilayer system by the adsorption of the aromatic molecule PTCDA.
  In this multilayer system, the interaction and corresponding charge transfer across the PTCDA/Pb interface quench the QWS in the Pb layer \cite{Stadtmuller.2019b}.
  This can potentially alter the momentum-dependent refilling processes within the Pb/Ag(111) bilayer system.

\begin{figure}
     \includegraphics[width=85mm]{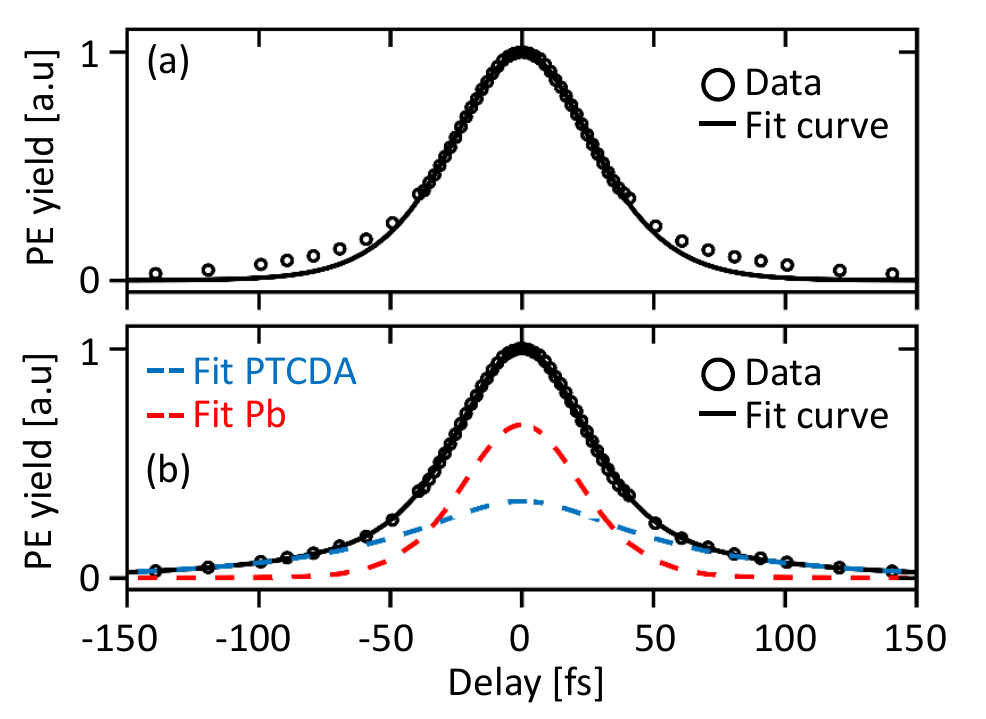}
  \caption{Exemplary normalized autocorrelation curves extracted from the k-space integrated photoemission yield of a time-resolved momentum microscopy experiment of the PTCDA/Pb/Ag(111) surface.
  As light source, a monochromatic laser setup with p-polarized light and 3.1 eV photon energy was used.
  The fits based on optical Bloch equations are plotted as black solid lines. The measured data in both panel is the same and presented as black circles.
  The data in panel (a) was fitted with one autocorrelation revealing a lifetime of $\tau=14.1\,\unit{fs}$ while the fit curve in region (b) is the sum of two separated autocorrelation curves, referring to the Pb and the PTCDA states.
  The lifetime value of the PTCDA fit $\tau_{\unit{PTCDA}}$ is $50.2\,\unit{fs}$ with a contribution of 33 \% to the summarized fit and the lifetime value of the Pb fit $\tau_{\unit{Pb}}$ is 8.3 fs with a weighting of 67 \% (A = 0.33).}
  \label{fig:Bild4}
  \end{figure}

  Upon the adsorption of PTCDA on Pb/Ag(111), the overall shape of the autocorrelation traces changes significantly.
  \mbox{Fig.\ref{fig:Bild4}(a)} shows an exemplary autocorrelation trace averaging the total momentum-resolved photoemission yield of the tr-2PMM experiment for a single intermediate state energy.
  The data was recorded with monochromatic radiation of $3.1\,$eV photon energy in gracing incidence geometry and p-polarized light pulses.
  The experimental data is illustrated as black circles.
  The best fitting result of this data with a single autocorrelation curve calculated by optical Bloch equations is included as a black solid line.
  This model curve can no longer describe the lineshape of the entire autocorrelation trace.
  In particular, we find clear deviations between the experimental data and the model curve for time delays between the pump and probe pulses larger than $\delta t\approx 50\,$fs.
  This underestimation of the experimental data by the modelled autocorrelation trace has already been observed for molecular adsorbates on surfaces and is a signature of an additional electronic state at this energy \cite{Steil.2013}.
  This second state must be an excited molecular state at this energy.

  To consider both states in our data analysis, we modelled the autocorrelation traces of the PTCDA/Pb/Ag(111) multilayer system with a linear combination of two autocorrelation traces exhibiting one lifetime $\tau_{\unit{PTCDA}}$ for the molecular adsorbate states and one lifetime $\tau_{\unit{Pb}}$ for the Pb-derived states.
  The relative contributions of both individual autocorrelation traces is modelled by weighting factors $A$ and $(1-A)$ respectively, with $A \in [0,1]$.
  The fitting result of this model is shown in \mbox{Fig.\ref{fig:Bild4}(b)} as black solid line.
  The individual contributions of the metallic and the molecular states are included as red and blue dashed curves respectively.
  The overall agreement of the fitting curve with the experimental data is excellent and we find a lifetime of $\tau_{\unit{Pb}} = 8.3\,$fs for the metallic state and of $\tau_{\unit{PTCDA}} = 50.2\,$fs for the molecular state.
  The weighting factor of the PTCDA contribution is $A=0.33$.

  The same model is used to determine the momentum-dependent lifetimes of the molecular and Pb-derived states.
  This approach results in individual CE lifetime maps for the molecular and the substrate states.
  To increase the reliability of our fitting procedure, we fixed the weighting factor $A$ for each energy to a constant value.
  The energy-dependent weighting factor was determined by analyzing the autocorrelation trace of the momentum integrated photoemission yield using $A$ as a free fitting parameter.

\begin{figure}
     \includegraphics[width=90mm]{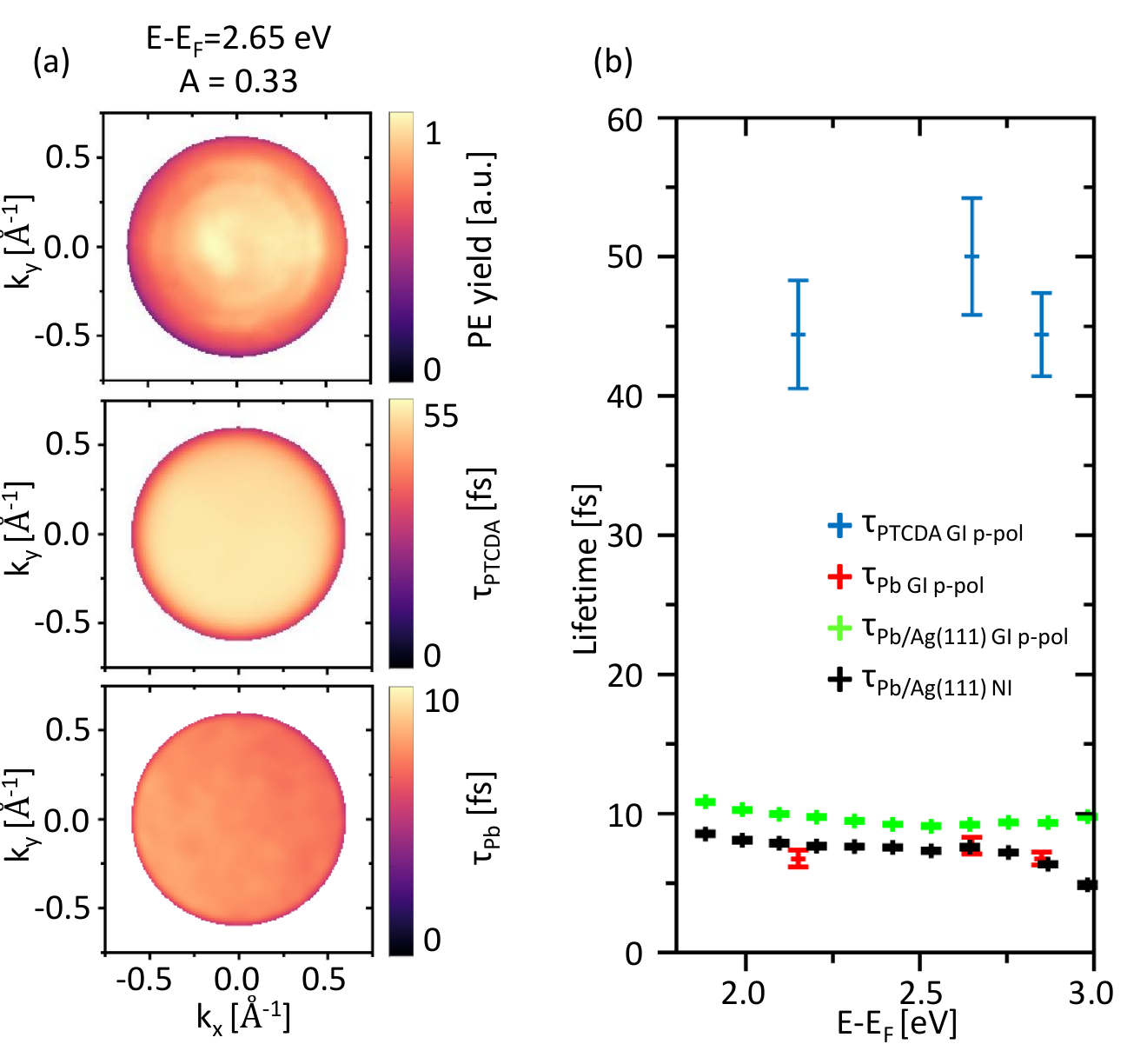}
  \caption{Time-resolved two-photon momentum microscopy experiments performed with p-polarized light, a gracing incidence angle of 65° and a photon energy of 3.1 eV. (a) Constant energy map and constant energy momentum-dependent lifetime maps of the PTCDA and Pb contributions for E-E$_{\unit{F}}=2.65\,$eV of the one layer PTCDA/Pb/Ag(111) system. (b) PTCDA (red) and Pb (blue) lifetimes extracted from a momentum-integrated autocorrelation trace in dependence of the intermediate state energy of the PTCDA/Pb/Ag(111) sample. The lifetimes for the Pb/Ag(111) system are depicted in green (p-pol, grazing incidence (GI)) and black (normal incidence (NI)) for comparison.}
  \label{fig:Bild5}
  \end{figure}

  Our findings for the momentum-dependent electron dynamics of the PTCDA/Pb/Ag(111) multilayer system is shown in \mbox{Fig.\ref{fig:Bild5}}.
  The monochromatic tr-2PMM data was obtained with p-polarized light, a gracing incidence angle of 65° and a photon energy of 3.1 eV.
  Constant energy maps of the momentum-dependent photoemission yield as well as the lifetime maps of the PTCDA- and Pb-derived states are exemplarily shown for an intermediate state energy of E-E$_{\unit{F}}=2.65\,$eV in \mbox{Fig.\ref{fig:Bild5}(a)}.
  The CE lifetime maps were obtained for a constant weighting factor $A=0.33$.

  At this energy, the CE intensity map reveals two ring-like features.
  The outer ring with azimuthal intensity modulation can be attributed to the Pb side band.
  The second ring with homogeneous azimuthal intensity distribution at smaller momenta is due to the Mahan cone transition.
  Despite these well-defined momentum space emission pattern, the CE lifetime maps of the molecular and Pb-derived states reveal homogeneous lifetimes for all momenta.
  This might not be surprising for the molecular states which only exhibit characteristic momentum space signatures for much larger momenta that are inaccessible in our tr-2PMM experiment \cite{Puschnig.2009,Stadtmuller.2012,Stadtmuller.2019b,Wallauer.2020}.
  However, we also do not observe any signature of a ring-like feature in the CE lifetime map of the Pb-derived side band as found for the CE lifetime map of the bare Pb/Ag(111) bilayer system for identical excitation conditions (p-polarized excitation).
  This directly points to a suppressing of any refilling process of electrons into the Pb side bands due to the adsorption of PTCDA.

Similar results were obtained for the intermediate state energies E-E$_{\unit{F}}=2.15\,\unit{eV}$ ($A=0.34$) and E-E$_{\unit{F}}=2.85\,\unit{eV}$ ($A=0.37$).
In particular, no signature of the band dispersion of the Pb side band can be observed in the CE lifetime maps of the Pb-derived bands for any energy.
These homogeneous lifetime distributions in all CE lifetime maps allow us to summarize the momentum-independent lifetime for the molecular and Pb-derived states in \mbox{Fig.\ref{fig:Bild5}(b)}.
The corresponding momentum-integrated lifetimes of the bare Pb/Ag(111) bilayer systems for excitation in normal and gracing incidence geometry are included as black and green dots respectively.

For the PTCDA/Pb/Ag(111) system, we find lifetimes of $\tau_{\unit{PTCDA}}\approx 50\,$fs for the molecular state and of $\tau_{\unit{Pb}}\approx 8\,$fs for the Pb-derived signals for all energies.
The lifetimes of the molecular state are comparable to previous studies of molecular materials on metallic surfaces \cite{Schwalb.2010}.
Interestingly, the lifetimes of the Pb-derived states reflect more closely the lifetimes determined for the bare Pb/Ag(111) bilayer system in normal incidence geometry, and not with those lifetimes determined in identical, gracing incidence geometry.

All these findings can be explained by the intrinsic hot electron dynamics of the bare Pb/Ag(111) bilayer system as well as by the modification of the Pb/Ag(111) band structure upon the adsorption of PTCDA.
For the bare Pb/Ag(111), optically excited electrons of the QWS are isotropically transferred into the Pb side band via interband scattering leading to the experimentally observed non-vanishing lifetime of the Pb side band with otherwise neglectable intrinsic quasiparticle lifetime.
Thus, the optically excited electrons in the QWS act as electron source for the refilling of the Pb side band.
The suppression of the Pb QWS upon the adsorption of PTCDA removes this source of electrons and hence suppresses the corresponding refilling of electrons into the Pb side band.

Interestingly, our experimental findings do not reveal any significantly transfer of electrons from molecular states into the Pb side band.
This suggests a marginal overlap between the p$_{\unit{x/y}}$-like orbitals of the Pb side bands with the molecular p$_{\unit{z}}$-like orbitals of the excited states.
This observation points to a rather weak, orbital-selective chemical interaction across the PTCDA/Pb interface.

\section{Summary}

  In our work, we have investigated the momentum-dependent electron dynamics of the quantum well system one layer Pb on Ag(111) prior and after the adsorption of PTCDA using tr-2PMM in different excitation geometries.
  The unoccupied band structure of the bare Pb/Ag(111) bilayer system is dominated by two excited states: (i) a free electron-like quantum well state with p$_{\unit{z}}$ orbital character and parabolic dispersion centered at the $\overline{\Gamma}$-point of the surface and (ii) a Pb side band with almost linear dispersion and p$_{\unit{x/y}}$ orbital character.
  The hot electron dynamics of electrons of the QWS is determined by isotropic intraband scattering leading to an increase of the lifetime from $12\,$fs to $13.2\,$fs from the highest accessible intermediate state energy towards the band bottom of the QWS.
  In addition, we find signatures of a (quasi-)elastic interband scattering process of electrons from the QWS into the Pb side band.
  This refilling process leads to a non-vanishing lifetime of the Pb side band with otherwise neglectable intrinsic quasiparticle lifetime.

After the adsorption of PTCDA on the Pb monolayer, the electron dynamics exhibits signatures of molecular and Pb-derived states, which both reveal a homogeneous electron dynamics in momentum space.
The lifetime of the molecular states is $\tau_{\unit{PTCDA}}\approx 50\,$fs for all energies.
The electron dynamics of the Pb-derived states is significantly altered by the adsorption induced suppression of the Pb QWS band structure.
This modification of the Pb band structure prevents any refilling of electrons within the Pb layer.
In addition, no interlayer refilling processes are observed which leads to vanishing lifetime of the Pb side band.

We attribute the adsorption-induced modification of the electron dynamics in the Pb layer to a weak, orbital-selective chemical interaction across the PTCDA/Pb interface.
This interaction selectively tunes the band structure and the corresponding intralayer scattering processes within the Pb quantum well systems.
In this way, our work lays the first steps towards controlling scattering processes of low dimensional systems and quantum (well) materials by the adsorption of organic molecules.

\begin{acknowledgements}

The experimental work was funded by the Deutsche Forschungsgemeinschaft (DFG, German Research Foundation) - TRR 173 – 268565370 Spin + X: spin in its collective environment (Project B05). BS and FH acknowledge financial support from the Graduate School of Excellence MAINZ (Excellence Initiative DFG/GSC 266). MC acknowledges funding from the European Research Council (ERC) under the European Union´s Horizon 2020 research and innovation programme (grant agreement No. 725767—hyControl).

\end{acknowledgements}


\end{document}